\documentclass[12pt,preprint]{aastex}

\newcommand{\kms}{km\ s$^{-1}$}

\newcommand{\Msun}{M$_{\sun}$}

\newcommand{\Ha}{H$\alpha$\ }
\newcommand{\bm}{beam$^{-1}$}
\newcommand{\yr}{yr$^{-1}$}

\slugcomment{ApJL, in press}

\shorttitle{Detection of CO in SMM J16359+6612}
\shortauthors{Sheth et al.}

\begin{document}

\title{Detection of CO from SMM J16359+6612, The Multiply Imaged Submillimeter
  Galaxy Behind A2218}

\author{Kartik Sheth\altaffilmark{1,2}, Andrew W. Blain\altaffilmark{1},
  Jean-Paul Kneib,\altaffilmark{1,3}, David T. Frayer\altaffilmark{4}, P. P. van der Werf\altaffilmark{5}, K. K. Knudsen\altaffilmark{5}}

\altaffiltext{1}{Division of Physics, Mathematics \& Astronomy, California
Institute of Technology, Pasadena, CA 91125}

\altaffiltext{3}{Laboratoire d'Astrophysique, OMP, 14 Av. E. Belin, 31499 Toulouse, France}

\altaffiltext{4}{Spitzer Science Center, California Institute of Technology, Pasadena, CA 91125}

\altaffiltext{5}{University of Leiden, Leiden, Netherlands}

\altaffiltext{2}{Email: kartik@astro.caltech.edu}

\begin{abstract}

We report the detection of CO ($J$=3$\rightarrow$2) line emission from
all three multiple images (A,B and C) of the intrinsically faint
($\simeq$ 0.8 mJy) submillimeter-selected galaxy SMM J16359+6612.  The
brightest source of the submm continuum emission (B) also corresponds
to the brightest CO emission, which is centered at $z$=2.5168,
consistent with the pre-existing redshift derived from \Ha.  The
observed CO flux in the A, B and C images is 1.2, 3.5 and 1.6 Jy
\kms\ respectively, with a linewidth of $500\pm 100$\,\kms.  After
correcting for the lensing amplification, the CO flux corresponds to a
molecular gas mass of $\sim 2\times 10 ^{10}\,h_{71}^{-2}$\,\Msun,
while the extent of the CO emission indicates that the dynamical mass
of the system $\sim$9$\times$10$^{10}$\,\Msun.  Two velocity
components are seen in the CO spectra; these could arise from either a
rotating compact ring or disk of gas, or merging substructure.  The
star formation rate in this galaxy was previously derived to be
$\sim$100--500 \Msun\ \yr.  If all the CO emission arises from the
inner few kpc of the galaxy and the galactic CO-to-H$_2$ conversion
factor holds, then the gas consumption timescale is a relatively short
40 Myr, and so the submm emission from SMM J16359+6612 may be produced
by a powerful, but short-lived circumnuclear starburst event in an
otherwise normal and representative high-redshift galaxy.

\end{abstract}

\section{Introduction}\label{intro}

When did the bulk of stars in galaxies form?  A huge number of studies
have sought to answer this question (e.g., \citealt{smail02,
giav04,steidel04} and references therein).  Observational evidence
indicates that the star formation activity and AGN activity peaked at
$z > $1, very likely between 2 $ < z < $ 3.  Sub-millimeter (submm)
surveys have shown that a large fraction of the total energy output of
galaxies in this critical epoch of galaxy formation comes from 
luminous, dusty galaxies \citep{smail97,blain99b}.  The most luminous
submm sources are widely believed to be the progenitors of today's
massive ellipticals \citep{lilly99,smail02}.  The slope of the submm
source counts rises steeply, indicating strong evolution, and the background 
radiation intensity at submm wavelengths is believed to be
dominated by sources with fluxes $\sim$1 mJy \citep{blain02}.  To
understand galaxy evolution, it is therefore imperative to study the
typically faint but numerous sub-mJy submm galaxies (SMGs).
Unfortunately, even with the best submm continuum data, these
intrinsically faint sources are difficult, if not impossible to study
at present.  The deepest existing 850\,$\mu$m continuum surveys are all
limited by confusion at 2 mJy \citep{blain02}.

Over 400 SMGs have been identified so far, yet their
distribution and evolution as a function of redshift has been hampered
by difficulties in identifying optical counterparts and obtaining
reliable redshifts until relatively recently
\citep{chapman03,chapman04}.  At the faint end of the luminosity
function (with intrinsic fluxes $<$ 10 mJy), only five
redshifts for submillimeter galaxies 
have been confirmed absolutely by the detection of CO spectral lines
\citep{frayer98,frayer99,neri03}.\footnote{A more substantial sample of 
SMGs has recently been compiled by Greve et al. (2004)} 
This paper presents the detection of CO
($J$=3$\rightarrow$2) emission using the Owens Valley Radio
Observatory's Millimeter Array (OVRO MMA) from the {\sl faintest}
submm galaxy to date, the remarkable object SMM J16359+6612
\citep{kneib04} that lies in a rare high-magnification configuration 
behind
the intervening massive cluster gravitational lens 
Abell 2218.  Its intrinsic
850$\mu$m continuum flux after correcting for the well-constrained
gravitational lensing effect of the foreground cluster is only
$\sim$0.8 mJy.

\section {OBSERVATIONS}

\begin{figure}
\plotone{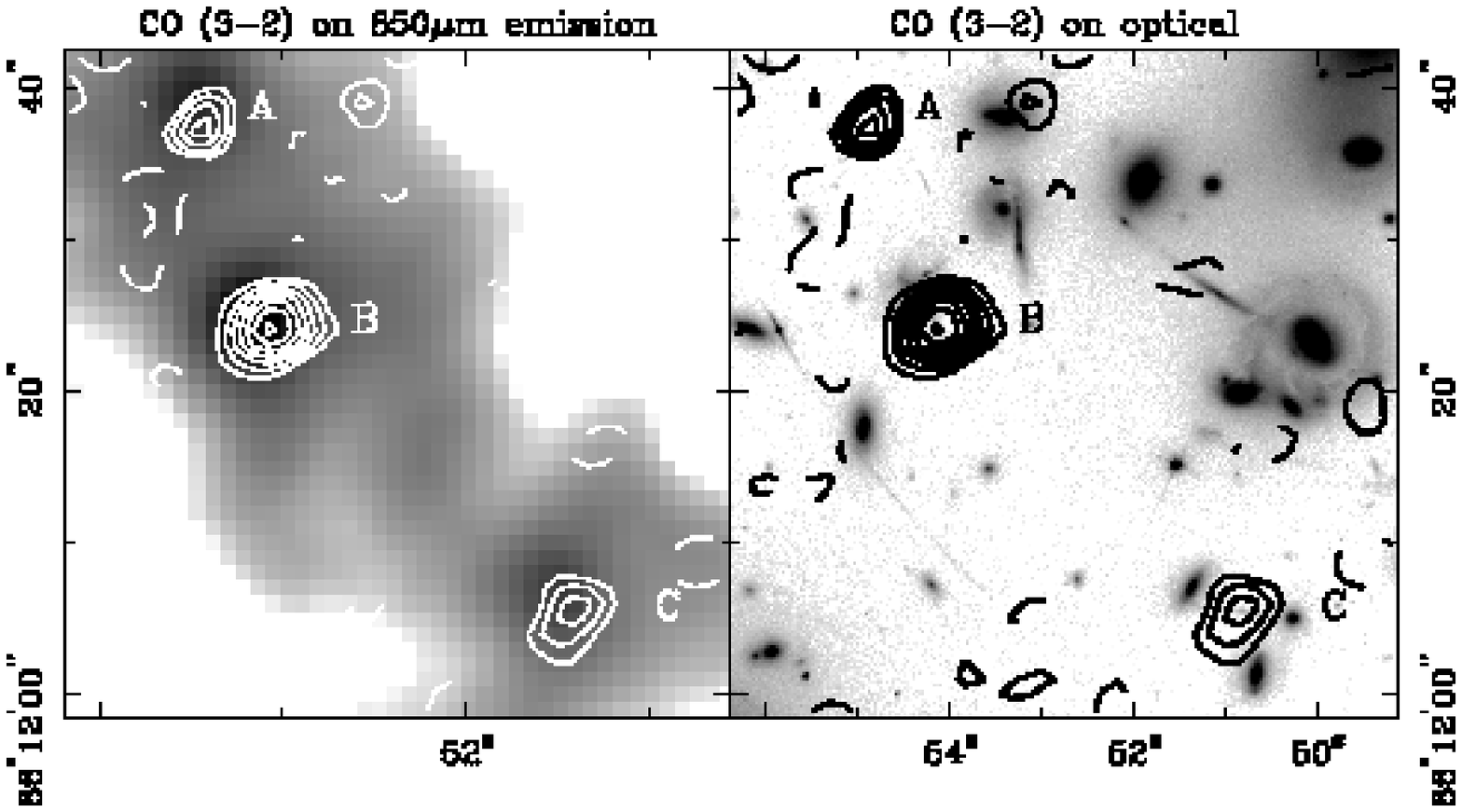}
\epsscale{0.9}
\caption{Contours show OVRO observations of the CO
  ($J$=3$\rightarrow$2) velocity-integrated, primary beam corrected,
  emission overlaid on both 850$\mu$m emission (left panel) and an
  optical {\it HST} image (right panel).  The three images are labeled
  A,B, and C following the convention of \citet{kneib04}.  The CO
  contour levels are $-$2,2,3,4,5... $\times$ 0.32 Jy \kms.  The peak
  of the 850$\micron$ submm continuum emission coincides with B at 17
  mJy.
 \label{fig1}}
\end{figure}

\begin{figure}
\plotone{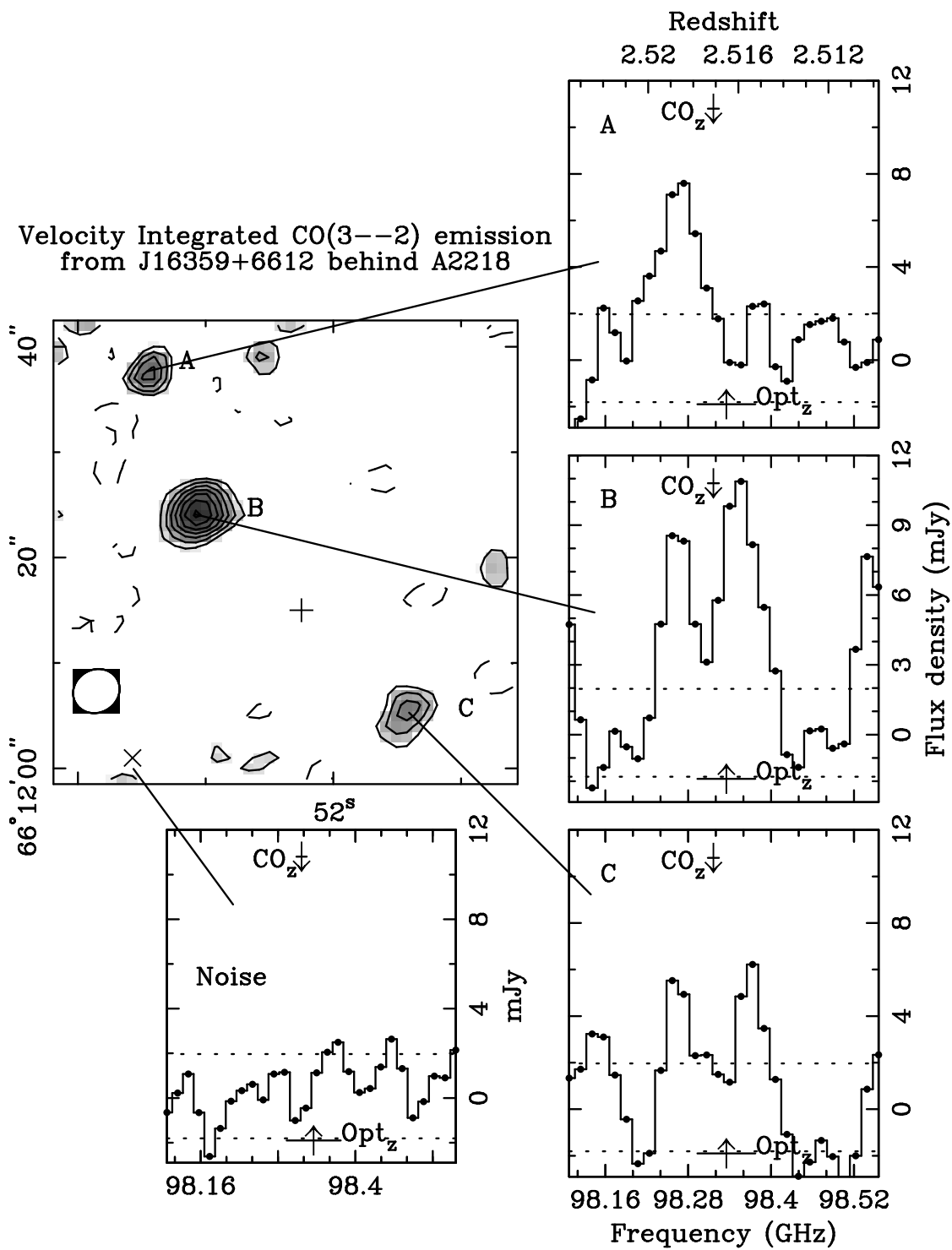}
\caption{A velocity-integrated, primary-beam corrected, CO
  ($J$=3$\rightarrow$2) emission map of SMM J16359+6612. The small
  plus sign between B and C indicates the OVRO phase and pointing
  center.  Hanning-smoothed spectra for the three lensed images are
  shown on the right and a noise spectrum is shown in a panel on the
  bottom left. The dashed line shows the 2$\sigma$ hanning smoothed
  noise level.  The CO and optically determined redshifts are indicated 
  with arrows.
\label{fig2}}
\end{figure}

We used OVRO MMA to search for CO ($J$=3$\rightarrow$2) line emission
from SMM J16359+6612 at the redshift ($z$=2.5165) derived from \Ha 
using Keck-NIRSPEC
\citep{kneib04} between October 2003 and April 2004 in three array
configurations, L, E and H.  The total integration time on source was
52.9 hours with a typical single-sideband temperature of 250K.  The
digital spectrometer was configured to observe at 98.3297 GHz with a
spectral bandwidth of 448 MHz and a spectral resolution of 4
MHz.  The phase and pointing center was $\alpha$(J2000) = 16:35:52.50
and $\delta$(J2000) = +66:12:15.00.  We used the nearby quasar
J1638+573 to correct for atmospheric phase and gain variations.  The
spectral bandpass was determined from observations of the quasars
3C84, 3C345 and/or 3c273.  Uranus and/or 3c273 were used for absolute
flux calibration, leading to a 15\% uncertainty in the overall flux
scale.  Data from each track was calibrated separately using the MMA
software package \citep{scoville93}, and the $uv$-data were imaged
using standard routines in the MIRIAD package \citep{sault95}.  The
resulting channel maps were binned to a spectral resolution of 16 MHz
(48.8 \kms), and have a synthesized beam of 4$\farcs$53 $\times$
4$\farcs$06 at a position angle of $-$59$^{0}$ with a noise level of
1.7 mJy \bm.  The channel maps were then corrected for primary beam
attenuation and combined into a velocity-integrated map (see Figure
\ref{fig1}).  In Figure \ref{fig2} we show the CO spectra corresponding
to each image, along with a typical representation of the noise.

\section {RESULTS}

The CO ($J$=3$\rightarrow$2) emission is brightest from the central
image.  Co-adding the spectra from the peak emission in each image and
fitting a single gaussian yields a redshift of z=2.5168$\pm$0.0003,
consistent with the \Ha redshift (z$_{H\alpha}$=2.5165$\pm$0.0015;
\citealt{kneib04}).  The CO kinematics are intriguing.  CO emission is
present at $>$ 2$\sigma$ over 400 \kms\ in two velocity components
located approximately $-$160 and +120 \kms\ with velocity widths of
$\Delta V_{\rm fwhm}\sim 200$\,\kms\ each.  The total CO linewidth
$\Delta V_{\rm fwhm} \sim 500 \pm 100$\,\kms. A consistent profile can
be seen in image C to the south.  The northern image A shows only the
red component of the CO emission but image A is at the edge of our
primary beam and suffers from a lower signal to noise ratio.  The CO
flux, corrected for primary beam attenuation, in the three images is
1.2$\pm$0.1, 3.5$\pm$0.1 and 1.6$\pm$0.1 Jy \kms, respectively, for a
total CO flux of 6.3 Jy$\pm$0.2 \kms.  To measure the molecular gas
mass in the system, we adopt $\alpha$ = 4 \Msun\ (K
\kms\ pc$^2$)$^{-1}$, similar to the Galactic value (see, e.g.,
\citealt{sanders91}), WMAP cosmology (H$_0$=71 \kms Mpc$^{-1}$,
$\Omega_M$=0.27, $\Omega_{\rm vac}$=0.73), and lensing amplification
numbers from \citet{kneib04}. Our choice of $\alpha$ = 4 \Msun\ (K
\kms\ pc$^2$)$^{-1}$ is consistent with that estimated for Arp 220
\citep{scoville97} after correcting for brightness ratio of T$_{b}$[CO
  ($J$=3$\rightarrow$2)]/T$_{b}$[CO ($J$=1$\rightarrow$0)] $\approxeq$
0.6 typically observed in starbursts \citep{devereux94}.  We follow
the formulae in \citet{solomon92} using a luminosity distance
calculated using WMAP cosmology.  We correct for the effect of
gravitational lensing by dividing the observed CO fluxes of each image
by its lensing amplification, and then compare the results inferred
from all three images.  Since A is at the edge of the primary beam and
suffers from lower signal to noise, we also give the masses derived
for each image separately (Table \ref{fluxtab}).  The total H$_2$ mass
is $\sim$ 2$\times$10$^{10}$ \Msun. Note that $\alpha$ is uncertain
and depends on metallicity and excitation; it may be as low as
$\alpha$ = 1 \Msun\ (K \kms pc$^2$)$^{-1}$ (e.g.,
\citealt{solomon97}), in which case the total H$_2$ mass is $\sim$ $5
\times 10^{9}$\,\Msun, comparable to the molecular gas mass in the
Milky Way.

We estimate a dust mass of $2\times10^{8}M_{\sun}$ from the 850$\mu$m
flux density, assuming a dust temperature of $T_{\rm d}=45$\,K.  The
inferred gas-to-dust ratio of 100 is consistent with metal-rich
galaxies in the local universe, suggesting that the central regions of
SMM\,J16359+6612 have already been significantly enriched with metals.

The CO emission from image B may be partially resolved in one
dimension.  Using the {\sl imfit} task in MIRIAD we fit the
velocity-integrated emission and measure a deconvolved size of
3$\farcs$0 $\times$ 1$\farcs$4 at a position angle of $-$68$^0$.
According to the \citet{kneib04} model, the linear stretch due to
lensing amplification along this position angle is about a factor of
2.  The major axis of the CO emission thus has an intrinsic extent of
about 12\,kpc, while the emission in the direction of the minor axis
is unresolved.  Note that the size estimates are upper limits given
the low signal to noise of the map; it is possible that the CO
emission is originating from a significantly smaller area.

\section{A Circumnuclear Starburst In Progress?}

SMM J16359+6612 has several remarkable properties.  The galaxy
morphology indicates an extremely red central core ($\gamma$ in
\citealt{kneib04}) bracketed by blue features ($\alpha, \beta$).
Unlike many submm galaxies, which show Ly$\alpha$ is present in
emission (e.g., \citealt{chapman03}), SMM J16359+6612 has Ly$\alpha$
in absorption, as in most Lyman-break galaxies (LBGs) 
\citep{shapley03}.  Given the complex morphology, the Ly$\alpha$ absorption,
the bright submm and infrared fluxes and 
compact \Ha emission, the star-formation activity in SMM\,J16359+6612 is
likely to be highly dust enshrouded.  The size of the 
red core in {\it HST} images suggests that the
star formation has an extent of only a few kpc \citep{kneib04}.  The
\Ha linewidth in this galaxy is only $\sim$280 \kms, which is greater than
for most LBGs (e.g., \citealt{pettini01,erb03}), but less than typically
found for high-redshift QSOs (e.g., \citealt{ivison98}).  
Nevertheless, the star
formation rate estimated from the far-infrared luminosity may be as high 
as 500\,\Msun\,\yr\ 
\citep{kneib04}, an order of magnitude
larger than in most LBGs (e.g., \citealt{pettini01,erb03}). Is SMM
J16359+6612 at a later evolutionary stage than submm sources
previously studied in CO?  Is it massive?  What will be the fate of
this galaxy and how long can it maintain its starburst/possible AGN
activity?

The molecular gas observations offer answers to some of these
questions.  The molecular mass inferred from the CO data
provides a modest reservoir of $\sim$2$\times$10$^{10}$ \Msun\ for
this rate of star formation activity.  The extent of the CO emission
indicates that the molecular gas is likely located in the central few
kpc, similar to the size of the red core in optical images
\citep{kneib04}.  If all the gas is consumed at the estimated current
star formation rate of 100--500 \Msun\ \yr, then the activity should
cease in a mere 40--200 Myr.  The nature of the star formation is
reminiscent of circumnuclear starbursts in local infrared
ultraluminous galaxies where the timescale for bursts is typically on
the order of $\sim$100 Myr, although the star formation rates and
central molecular concentration in the lower redshift galaxies are
both less by about a factor of 10.  The gas and star formation
activity in local starbursts are often distributed in a ring-like or
disk-like morphology (e.g., \citealt{downes98}).  In this respect too
SMM J16359+6612 could resemble a circumnuclear starburst: the two CO
velocity components may indicate a gas ring or disk morphology.  If
the CO were centrally concentrated (and in the absence of merging) we
would expect a single broad velocity peak in the CO emission.

There is one troubling discrepancy between the CO and \Ha
observations.  The CO linewidth is about 500 \kms\ whereas the \Ha
linewidth is 280 \kms.  One explanation for this difference is that
the two lines trace different parts of the galactic potential: the \Ha
emission is likely confined to only the central 1--2 kpc in the source
plane \citep{kneib04} whereas the CO emission may come from a more
extended region.  This distribution could be consistent with a ring
morphology for the CO distribution if the central molecular gas has
been consumed by the starburst activity.  Another possibility is that
we are only observing the \Ha emission from the near side of the
galaxy or through windows where there is less extinction from dust.  
As the CO
emission is only barely resolved, higher spatial resolution
observations (e.g., with IRAM, CARMA\footnote{The Combined Array for
  Research in Millimeter-wave Astronomy, http://www.mmarray.org} and
eventually ALMA\footnote{Atacama Large Millimeter Array,
  http://alma.nrao.edu}) are needed to understand the relative
distribution of the molecular gas and star formation activity.

\section{Is This Submm Source Fated to Become a Massive Early Type Galaxy?}

Do all galaxies pass through a bright-submm phase?  This is an
outstanding question in the literature (e.g.,
\citealt{lilly99,smail02}).  Certainly the brightest submm sources at 
near-IR wavelengths are
believed to be analogs of today's local, massive, early type galaxies
\citep{smail02, smail04}.  But what about the fainter submm sources that
contribute the majority of the submm background?  Do later Hubble
types and less massive galaxies also pass through a bright submm phase
when building their bulges and perhaps when their stellar disks begin
to form?  If so, then the corresponding fainter submm galaxies 
should not be very
massive ($<10^{12}$\,\Msun).  With the CO data for SMM J16359+6612, 
comparable to that available for cB58 \citep{baker04}, we
can make an attempt to address this fundamental question.  

Assuming that the molecular gas is tracing the galaxy mass at a radius
of 6 kpc, and using the CO linewidth of 500 \kms, we estimate a
dynamical mass of 9$\times$10$^{10}$ \Msun. Note, however, that the
mass is uncertain by at least a factor of two and depends on several
important assumptions.  The mass may be smaller depending on the true 
size of the CO emitting region which is, at best, only barely
resolved in these data.  The mass could also be higher depending on the
inclination of any disk present.

Not surprisingly, the dynamical mass in SMM J16359+6612 is at the low
end of SMGs (see review by \citealt{genzel04} and references therein),
and is higher than the typical dynamical masses ($\sim$10$^{10}$\Msun)
inferred for LBGs \citep{pettini01,erb03,baker04,baker04b}.  But unlike
previous SMGs, and especially the strongly lensed LBG cB58 for 
which CO spectroscopy is available \citep{baker04}, 
SMM J16359+6612 lacks a large molecular gas reservoir by fraction. 
The relatively low gas fraction suggests that the bulk of the stellar mass in
this system is already in place and so the stellar mass is unlikely to
increase significantly without subsequent merging activity.  So does
this imply that a submm-bright phase is typical for all galaxies
including the less massive galaxies in the field today?

SMM J16359+6612 is part of a close galaxy
group: two other galaxies (\#384/\#478 and \#273), neither of which
are detected in the submm, are within 130 kpc, and all three are
separated by less than 100 \kms\ from each other in redshift.  The
submm luminosity of SMM J16359+6612 may be boosted by star-formation
triggered by the interaction with these neighbors 
\citep{kneib04}.  This situation is
analogous to local galaxies where otherwise normal spirals undergo
massive bursts of star formation and molecular gas inflow upon an
interaction or merger (e.g., \citealt{scoville97} and references
therein).  We cannot conclusively state that the faint end of the
submm population is composed exclusively of low mass systems. However, 
as SMM J16359+6612 appears to be less massive than the brightest
submm sources, it 
is clear that a fraction of faint submm galaxies 
could be systems with low gas fractions and more modest total masses 
that could have elevated submm
fluxes due to intense but short-lived starburst activity.

\section{Conclusions}

Our detection of molecular gas from SMM J16359+6612, the faintest
submm-selected galaxy observed in CO emission to date, reveals some remarkable
properties.  SMM J16359+6612 is significantly
smaller in extent, and has a 
lower dynamical mass {\sl and} a lower gas mass fraction than previous 
SMGs. These factors all suggest that 
SMM J16359+6612 is unlikely to develop into a massive elliptical galaxy 
without extensive subsequent merging.  
The distribution and kinematics of the
molecular gas and star formation activity indicate that a circumnuclear
starburst is likely to be responsible for the submm emission, in
analogy with nearby ultraluminous infrared galaxies. The luminous activity
is likely induced by interactions with two neighboring galaxies.
SMM J16359+6612 indicates that
at least a fraction of the sub-mJy submm sources are likely going through
a bright submm phase due to an episode of intense star formation activity 
that is found in a galaxy of relatively modest mass, and triggered by 
environmental factors. This type of interaction-induced activity in 
a modestly massive system could 
help to explain the very strong clustering signal seen in submm 
galaxies \cite{blain04}.
 
\acknowledgments

We thank the anonymous referee for many helpful comments that helped
strengthen this paper.  We thank our colleagues at the Owens Valley
Millimeter Array who have helped make these observations possible.  We
are also thankful to Karin Menendez-Delmestre, Nick Scoville, Alexie
Leauthaud and Dawn Erb for their comments and helpful suggestions.
The Owens Valley Millimeter Array is operated by the California
Institute of Technology and is partially supported by NSF grant
AST-9981546 and the Norris Foundation.  AWB acknowledges support from
the NSF under grant AST-0205937, the Alfred P. Sloan Foundation, and
the Research Corporation. JPK acknowledges support from CNRS and
Caltech.

\clearpage
%
%

\clearpage

\clearpage

%
%
%
\begin{deluxetable}{crrccrr}
\tablecaption{OVRO CO(3 --2) Data for the three images of SMM J16359+6612\label{fluxtab}}
\tablehead{\colhead{ID} &\colhead{$\alpha$(J2000)} &\colhead{$\delta$(J2000)} &\colhead{S$_{CO}$} &\colhead{Amplification} &\colhead{Intrinsic S$_{CO}$}&\colhead{M(H$_2$)} \\
\omit & \omit & \omit & \colhead{(Jy \kms)} & \omit &\colhead{(Jy \kms)}&\colhead{(10$^{10}$ \Msun)}}
\startdata
A & 16$^h$35$^m$54.814$^s$ & 66$^0$12'37'' & 1.2$\pm$0.14$^a$ & 14 & 0.09$^a$ & 1.3 \\
B & 16$^h$35$^m$54.152$^s$ & 66$^0$12'24'' & 3.5$\pm$0.12 & 22 & 0.16 & 2.3 \\
C & 16$^h$35$^m$50.848$^s$ & 66$^0$12'06'' & 1.6$\pm$0.13 & 9 & 0.18 & 2.6 \\ 

\enddata
\tablenotetext{a}{Image A is at the edge of the primary beam and suffers from low signal-to-noise ratio, as noted in the text.} 
\end{deluxetable}
\end{document}